\documentstyle[11pt,aaspp]{article}
\lefthead{GARNETT ET AL.}
\righthead{CARBON IN DWARF GALAXIES}
\slugcomment{To appear in the 10 April 1995 {\it ApJ} }
\begin{document}
\title{THE EVOLUTION OF C/O IN DWARF GALAXIES FROM HST FOS
OBSERVATIONS\altaffilmark{1}}
\author{Donald R. Garnett\altaffilmark{2} and Evan D. Skillman}
\affil{Astronomy Department, University of Minnesota\\ 116 Church St., S. E.,
Minneapolis, MN 55455 \\ garnett@oldstyle.spa.umn.edu,
skillman@zon.spa.umn.edu}
\author{Reginald J. Dufour}
\affil{Department of Space Physics and Astronomy, Rice University, Houston, TX
77251-1892 }
\author{Manuel Peimbert and Silvia Torres-Peimbert}
\affil{Instituto de Astronomia, UNAM, Apartado Postal 70-264, DF 04510 Mexico,
Mexico }
\author{Roberto Terlevich and Elena Terlevich}
\affil{Royal Greenwich Observatory, Madingley Road, Cambridge CB30EZ, England}
\and
\author{Gregory A. Shields}
\affil{Astronomy Department, University of Texas, Austin, TX 78712}
\altaffiltext{1}{Based on observations with the NASA/ESA Hubble Space
Telescope obtained at the Space Telescope Science Institute, which is
operated by Association of Universities for Research in Astronomy,
Incorporated,
under NASA contract NAS5-26555.}
\altaffiltext{2}{Hubble Fellow.}

\vfill\eject
\begin{abstract}

We present UV observations of seven H~II regions in low-luminosity dwarf
irregular galaxies and the Magellanic Clouds obtained with the Faint Object
Spectrograph on the Hubble Space Telescope in order to measure the C/O
abundance ratio in the ISM of those galaxies. We measure both O~III] 1666~\AA\
and C~III] 1909~\AA\ in our spectra, enabling us to determine
C$^{+2}$/O$^{+2}$ with relatively small uncertainties. The results from our
HST observations show a continuous increase in C/O with increasing O/H,
consistent with a power law having an index of 0.43$\pm$0.09 over the
range $-$4.7 to $-$3.6 in log(O/H). One possible
interpretation of this trend is that the most metal-poor galaxies are the
youngest and dominated by the products of early enrichment by massive stars,
while more metal-rich galaxies show increasing, delayed contributions of carbon
from intermediate mass stars. However, recent evolution models for massive
stars including mass loss suggest that the yield of carbon from massive stars
may increase with metallicity relative to the yield of oxygen; new chemical
evolution models for the solar neighborhood which include nucleosynthesis
from these recent stellar evolution models predict a C/O abundance evolution
similar to that observed in the metal-poor galaxies. The trend in the C/N ratio
increases steadily with O/H in the irregular galaxies, but decreases suddenly
for solar neighborhood stars and H II regions. This may indicate that the
bulk of nitrogen production is decoupled from the synthesis of carbon in
our Galaxy. Our results also suggest that it may not be appropriate to combine
abundances in irregular galaxies with those in spiral galaxies to study the
evolution of chemical abundances. Our measured C/O ratios in the most
metal-poor
galaxies are consistent with predictions of nucleosynthesis from massive stars
for Weaver and Woosley's best estimate for the $^{12}C(\alpha,\gamma)^{16}O$
nuclear reaction rate, assuming negligible contamination from carbon produced
in intermediate mass stars in these galaxies.

We detect a weak N~III] 1750~\AA\ multiplet in SMC N88A and obtain interesting
upper limits for two other objects. Our 2$\sigma$ upper limits on the 1750~\AA\
feature indicate that the N$^{+2}$/O$^{+2}$ ratios in these objects are not
significantly larger than the N$^+$/O$^+$ ratios measured from optical spectra.
This behavior is consistent with predictions of photoionization models,
although better detections of N~III] are needed to confirm the results.
\end{abstract}
\section{Introduction}

The evolution of the relative abundances of the chemical elements provides
clues
to the star formation history and evolution of a galaxy (Wheeler, Sneden, \&
Truran 1990). Carbon and oxygen are among the most important in this regard,
and knowledge of the variation of their abundances are important for a variety
of problems, including:
(1) C and O constitute the bulk of the matter that is not hydrogen or helium,
and thus are important sources of interior opacity in stars. Knowledge of the
time evolution of C and O abundances is necessary to properly model the
structure and evolution of stars at different ages; e.g., theoretical
isochrones
can give different ages for globular clusters for different values of O/Fe in
the stars (VandenBerg 1985).
(2) The formation of the CO molecule in the ISM should be regulated by the
relative abundances of C and O at some level, so an understanding of the
variation of C/H and C/O in the ISM is relevant for modeling the formation of
CO molecules and the possible effects of abundance variations on the
I(CO)/N(H$_2$) relation.
(3) There is evidence that the interstellar dust-to-gas ratio is related to
the interstellar carbon abundance (Mathis 1990), so knowledge of C (and also
Si) abundances in a variety of environments is relevant to understanding the
amount, composition, and evolution of dust in those environments.
(4) In theory, oxygen is synthesized almost entirely in massive stars
(M $>$ 10~M$_\odot$ ), while carbon is produced in both massive and
intermediate
mass stars. Thus, the ejection of some carbon is delayed in time with respect
to oxygen, so the C/O ratio offers a potential ``clock'' for determining the
relative ages of stellar systems.
(5) Steigman, Gallagher, \& Schramm (1989) have argued that the C abundance
is a better measure of stellar He production than O, and therefore that it
is better to compare He/H with C/H when extrapolating to the primordial He
abundance. Few observations of C abundances are available to test this
controversial hypothesis.

The primary obstacle to our understanding the evolution of carbon in other
galaxies is the severe lack of high-quality measurements of carbon abundances.
C has no strong, easily observable transitions in the optical; the most
important species, C III, has emission lines of reasonable strength only
in the UV (C III] 1907, 1909\AA), requiring spacecraft observations, while
[C II] only has observable transitions in the UV (2324-2329\AA) and far
infrared. Considerable observational effort has been made attempting to
measure C abundances and C/O ratios within H II regions in our own and
other galaxies, from ground-based optical spectra and UV spectra from IUE
(a brief list of references includes Torres-Peimbert et al. 1980; Peimbert
et al. 1986; Dufour et al. 1984, 1988; Skillman 1990, Walter et al. 1992,
and Peimbert, Torres-Peimbert \& Dufour 1993). Useful observations of C
lines in about a dozen
H II regions have been obtained with IUE, but for most of these the C line
detections are marginal. There are also large uncertainties associated with
relating the UV spectrum to the optical lines due to corrections for reddening
and for differences in aperture sizes for the optical and the UV observations
(Dufour \& Hester 1990). Hence, a typical carbon abundance measurement from
IUE observations is uncertain by at least a factor of 2-3. In nearly all cases,
improved measurements of the C lines are desirable.

We can overcome many of these uncertainties by observing collisionally-excited
UV emission lines from C$^{+2}$ and O$^{+2}$ simultaneously. IUE did not have
the sensitivity necessary to measure O~III] $\lambda$1666 except in
high-excitation planetary nebulae. Therefore, we began a program of
observations
with the FOS on the HST of the intercombination doublets O III] 1661-1666\AA\
and C III] 1906-1909\AA\ in giant H II regions in a sample of
dwarf emission-line galaxies to determine the general behavior of the C/O
ratio at low metallicities. We report on the results of that program here.


\section{Observations and Analysis}

\subsection{HST Observations}

We obtained UV spectra of metal-poor dwarf emission-line galaxies with the
FOS on HST during Cycles 2 and 3. A journal of the observations and the HST
observation identifications are given in Table 1. Most of the spectra
were taken with grating G190H and the 1$^{\prime\prime}$ circular aperture
(aperture B-3) to achieve 3 \AA\ spectral resolution over the wavelength range
1600-2300 \AA. This range includes important nebular emission features from
C$^{+2}$ (1907, 1909 \AA) and O$^{+2}$ (1661, 1666 \AA), as well as N$^{+2}$
(1748-1754 \AA) and Si$^{+2}$ (1882, 1892 \AA). Our integrations were chosen
to be long enough to detect both the O~III] and C~III] features to 4$\sigma$
or better; C~III] is typically much stronger than O~III] in H~II regions, so
the S/N in O~III] is the limiting factor in the accuracy of our results. This
observational approach has many advantages over previous studies which combined
UV data from IUE with ground-based data:

-- measuring O~III] and C~III] together eliminates observational difficulties
in combining ground- and space-based spectra, such as mismatched apertures and
positioning uncertainties;

-- the interstellar extinction curve is nearly flat over the short wavelength
range 1600--2000 \AA. Thus, the relative line strengths suffer very little
from uncertainties due to reddening;

-- the UV O~III] and C~III] lines have similar excitation potentials, so that
uncertainties in T$_e$ have only a minor effect on the derived ionic
abundances;

-- O$^{+2}$ and C$^{+2}$ also have similar ionization potentials, and so
corrections for nebular ionization structure are also relatively small.

One observation, that of NGC 2363, was taken with G190H as well, but through
the 1$^{\prime\prime}$ square paired aperture (aperture C-1). This sampled
two regions of the nebula separated by 3$^{\prime\prime}$. Only one of the
positions had sufficient signal/noise to be useful for our abundance
determinations, and we present results for that position only. Finally,
for the H~II region N88A in the SMC we use a spectrum taken through the
0.7$^{\prime\prime}$x2.0$^{\prime\prime}$-BAR aperture (aperture C-4),
centered on the ionizing star.

The target objects (typically bright H~II regions located in low surface
brightness galaxies) were positioned in the spectrograph aperture by first
centering on a nearby star (less than one arcminute away) using a binary
acquisition, then transferring to the target via a blind offset. Following
each spectroscopic observation we obtained a short white-light image through
the 4.3$^{\prime\prime}$ target acquisition aperture to check the positioning.
In two cases (I~Zw~18 and T1214-277) the aperture missed the strong emission
line peak by approximately one arcsecond and sampled a
lower surface brightness region instead, resulting in lower signal to
noise in those spectra. We did obtain a weak detection of O~III] in T1214-277,
but only an upper limit for O~III] in I Zw 18. For all of the observations for
the other targets, the acquisition images showed that the spectrograph aperture
was at the correct position.

The spectra were processed through the standard HST pipeline reductions,
with corrections for flat-fielding, wavelength determination, instrumental
background subtraction, and photometric calibration. The instrumental and
sky backgrounds are negligible for our observations; our objects are bright
in the UV compared to the background, and there
are no significant geocoronal emission lines in the wavelength region of
interest. For the NGC 2363 observation we suppressed the sky subtraction
option and simply summed the total spectrum at each position. There is no
direct photometric calibration for the C-1 aperture. However, our scientific
analysis requires accurate relative spectrophotometry only, and since the
different FOS apertures should not introduce systematic color errors in the
flux calibration, we used the (then current) calibration for the B-3 aperture
scaled by the difference in aperture size. As a result, the absolute fluxes
for NGC~2363 may not be accurate. Examples of our FOS spectra are shown in
Figure 1.

Line fluxes were measured by direct integration under the line profiles and
by fitting Gaussian profiles to the lines. Both methods gave very good
agreement for lines with high signal/noise. Our measured line fluxes
are listed in Table 2. The uncertainties in the line fluxes were determined
from the statistical noise over the wavelength interval of each line
(determined by the FWZI) combined with the uncertainty in the relative
spectrophotometry of the FOS, which is less than 5\% (Bohlin, private
communication). Our 1$\sigma$ uncertainties in the line strengths computed
in this way are listed in Table 2.

\subsection{Supporting Optical Observations}

For our targets there exist excellent ground-based optical spectra, either
published or in preparation for publication. The optical spectra can be
found in Campbell, Terlevich \& Melnick (1986) for C1543+091; Pagel et
al. (1992) for T1214-277; Skillman \& Kennicutt (1993) for I~Zw~18;
Gonzalez-Delgado et al. (1994) for NGC~2363; Mathis, Chu \& Peterson
(1983) for 30~Doradus; Terlevich et al. (in preparation) for SBS0335-052;
and Dufour et al. (in preparation) for N88A. From the optical spectra we
obtain physical parameters needed to complete the analysis of the UV
spectra: interstellar reddening, electron temperature T$_e$, electron
density n$_e$, abundances for oxygen and other heavy elements, and the
degree of ionization of the nebular gas, measured by O$^{+2}$/O.

We use the normal interstellar extinction law of Seaton (1979) to estimate
the reddening correction based upon the measured extinction values from our
optical spectra. The difference in
reddening between 1650 \AA\ and 1910 \AA\ is only 0.023 dex for A$_V$ = 1
magnitude, or less than 5\%. Except for 30~Doradus, our estimated reddenings
are well under 1 magnitude visual, and so interstellar reddening does not
seriously affect the observed relative strengths of C~III] and O~III]. For
N88A the bulk of the observed extinction is local to the SMC, and in this
case we use the SMC extinction law of Prevot et al. (1984) to correct the
relative strengths of the UV lines. The corrected UV line strengths normalized
to C~III] are listed in Table 2.

\section{Ionic and Elemental Abundances}

\subsection{Physical Conditions}

{}From the optical measurements listed in \S 2.2 we obtain estimates of
the electron temperatures (from the [O~III] lines) and electron densities
(from the [S~II] lines) in our FOS target H~II regions. The measured
electron temperatures are listed in Table 3; the densities in these
objects ($<$ 10$^3$ cm$^{-3}$) are too small for significant deactivation
of the UV transitions we observe. Our FOS position in 30~Doradus is
located on the bright knot between positions III-2 and III-3 from Mathis
et al. (1983); we used an average of the derived physical parameters and
abundances at those two positions to analyze our FOS spectrum.

\subsection{Ionic Abundances}

Our analysis is relatively straightforward. The electron densities for
our H~II regions are all well below the critical densities for collisional
de-excitation of the C~III] and O~III] intercombination lines. Therefore,
we can compute the C$^{+2}$/O$^{+2}$ abundance ratio in the low density
limit, as demonstrated in Aller (1984), chapter 5, and Osterbrock (1989),
chapter 5. In the low density limit, the emission rate coefficient for a
collisionally-excited emission line is given by
\begin{equation}
j(\lambda) = h\nu{8.629\times10^{-6}\over{T_e^{0.5}}}{\Omega(1,2)\over
{\omega_1}}
exp(-\chi/kT_e)N_eN(X^{+i}),
\end{equation}
where T$_e$ and N$_e$ are the electron temperature and density, $\Omega$(1,2)
is the effective collision strength between the two levels, $\omega_1$ is the
statistical weight of the lower level, $\chi$ is the excitation potential of
the transition, and N(X$^{+i}$) is the number density of the ion under
consideration.

The main useful sources of effective collision strengths for C~III
$^3$P-$^1$S$^o$ at electron temperatures below 20,000 K are Mendoza (1983)
(based upon the computations of Dufton et al. (1978) and Berrington et al.
(1977) plus data from the Daresbury atomic database), and Berrington (1985).
Berrington does not provide collision strengths below log T$_e$ = 4.1, but
the collision strength for $^3$P-$^1$S$^o$ is nearly constant between 10,000 K
and 20,000 K, suggesting reliable extrapolation is possible. Mendoza's
collision strengths are only 4-5\% smaller than Berrington's, and here we
adopt the mean values from the two papers. Nevertheless, new calculations
of C III collision strengths over the full range of nebular temperature
conditions are desirable.

For O~III $^5$S$^o$-$^3$P (from which the $\lambda$1661, 1666 transitions
arise)
there are several similar, independent calculations (Baluja, Burke, \& Kingston
1981 (BBK), Aggarwal 1983 (A83), and Burke, Lennon, \& Seaton 1989 (BLS)).
Comparing the three calculations, we see that BLS and A83 find similar
values for $\Omega$
as a function of T$_e$, while the values from BBK are smaller by 6-8\%. On the
other hand, for $\Omega$($^1$D-$^3$P) BLS obtain values higher than BBK and A83
by about the same amount, while for $^1$S-$^1$D the three calculations give
$\Omega$'s spread out approximately evenly over a range of about 7\%. It may
be that this range reflects the intrinsic precision of the current electron
impact calculations. Since this spread is small compared with other
uncertainties
in our analysis, we take the mean of the collision strength values from BBK,
A83, and BLS for our abundance calculations, with an adopted uncertainty of
$\pm$5\%. The collision strength values we adopt for three different
temperatures
are listed in Table 4.

With these values for the C~III and O~III collision strengths the
C$^{+2}$/O$^{+2}$ ionic abundance ratio can be computed directly from
the C~III]/O~III] line ratio by using equation (1) to derive the relation
\begin{equation}
{C^{+2}\over O^{+2}} = 0.089~e^{-1.09/t}~{I(\lambda 1909)\over I(\lambda
1666)},
\end{equation}
where $t = T_e/10^4 K$ and $I(\lambda)$ is the line intensity. The numerical
coefficient before the exponential has a small temperature dependence, varying
by 6\% over the range 10,000 - 20,000 K, with the average value shown in
equation
(2). All of the H~II regions in our
sample have electron temperatures within this range. Our computed
C$^{+2}$/O$^{+2}$ ratios are listed in Table 3. In these calculations we
have assumed that our measured [O~III] electron temperatures are
appropriate for both C~III] and O~III].
This is a reasonable assumption given that the C~III] and O~III] lines are
similar in both ionization and excitation. In addition, Garnett (1992) compared
characteristic ion-weighted electron temperatures for various ions using
photoionization modeling, and found little systematic variation of T(C~III)
with respect to T(O~III).

Despite the lack of detections of the N~III] $\lambda$1750 multiplet, we can
estimate upper limits on N$^{+2}$/O$^{+2}$ in a similar way to
C$^{+2}$/O$^{+2}$.
The relation is
\begin{equation}
{N^{+2}\over O^{+2}} = 0.212~e^{-0.43/t}~{I(\lambda 1750)\over I(\lambda
1666)},
\end{equation}
using the collision strengths for N~III computed very recently by Blum \&
Pradhan
(1992). Our 2$\sigma$ upper limits, determined over a 10 \AA\ bandpass, are
listed
in Table 3. We did obtain a weak detection of N~III] in N88A, and we list the
computed N$^{+2}$/O$^{+2}$ ratio based on that detection in Table 3.


\subsection{Corrections for Ionization}

To first order, one can say that C$^{+2}$ and O$^{+2}$ are similar ionization
states, and therefore that C$^{+2}$/O$^{+2}$ $\approx$ C/O. However, the
various
ions of C and O have different ionization potentials: for C$^+$ and C$^{+2}$,
the IPs are 24.4 eV and 47.9 eV, respectively, while for O$^+$ and O$^{+2}$ the
IPs are 35.1 and 54.9 eV. Thus, we might expect non-negligible amounts of
C$^{+3}$ in H~II regions ionized by the hottest O stars, while in nebulae
ionized by cooler stars the fraction of carbon in C$^{+2}$ may exceed the
O$^{+2}$
fraction. The relative ionization of these species can vary systematically if,
for
example, stellar ionizing continua vary with metallicity, the stellar
mass-T$_{eff}$
relation varies with metallicity (Maeder 1990), or if the stellar mass function
varies with metallicity (Terlevich \& Melnick 1985). Therefore, we must look at
photoionization models to estimate the corrections for ionization to convert
our
measured C$^{+2}$/O$^{+2}$ ratios to true C/O abundance ratios.

To look at this we have computed photoionization models using the code
described
in Shields et al. (1981) and Garnett (1989). The model nebulae were computed as
dust-free spherical nebulae consisting of filaments in a vacuum with some
filling
factor $\epsilon$. We constructed ionizing continua for OB associations in a
manner
similar to that employed by McGaugh (1991). We used stellar mass - effective
temperature relations at metallicities Z = 0.001, 0.004, and 0.008 from the
stellar
evolution models of Schaller et al. (1992), Schaerer et al. (1993a,b), and
Charbonnel
et al. (1993) for both
ZAMS and 2 Myr old stars. Lyman-continuum luminosities as a function of stellar
mass were then determined from the T$_{eff}$ - N(Ly-c) relation of Panagia
(1973)
for the ZAMS; for 2 Myr old stars, we scaled N(Ly-c) from the ZAMS value as the
bolometric luminosity. A Salpeter initial mass function over the range 10 - 100
solar masses was used to determine the relative contribution of a given stellar
mass to the total ionizing luminosity of the cluster. Stellar fluxes from the
non-LTE models of Mihalas (1972) determined the shape of the ionizing continua.

The nebula models were computed for the same metallicities as the stellar
evolution models, i.e., Z = 0.001, 0.004, and 0.008. The models included the
elements H, He, C, N, O, Ne, Mg, Si, S, and Ar. With the following exceptions,
the elements heavier than H were kept in their solar system ratios:

(1) we took N/O = 0.03, the average value observed in metal-poor irregular
galaxies (Garnett 1990);

(2) Mg/O and Si/O were reduced by a factor of 10 from their solar system ratios
to simulate depletion of Mg and Si onto grains.

(3) The He mass fraction Y was varied as Y = 0.23 + 5$\Delta$(O/H).

In computing the models the number of ionizing photons was kept fixed at
log N(Ly-c) = 52.0 photons s$^{-1}$ while the ionization parameter U (as
defined in Shields \& Searle 1978) was varied by adjusting the gas filling
factor. The models cover the range $-$3.0 $<$ log U $<$ $-$2.0, the typical
range for giant H~II regions (Shields 1990).

The results from the models are displayed in Figure 2, where we plot the
ratio of the C$^{+2}$ and O$^{+2}$ volume fractions, X(C$^{+2}$)/X(O$^{+2}$),
versus the O$^{+2}$ fraction X(O$^{+2}$). In terms of the ionic abundance ratio
C$^{+2}$/O$^{+2}$, the elemental abundance ratio C/O is then
\begin{eqnarray}
{C\over O} & = & {C^{+2}\over O^{+2}}\Biggl[{X(C^{+2})\over
X(O^{+2})}\Biggr]^{-1}\\
{        } & = & {C^{+2}\over O^{+2}} \times ICF, \nonumber
\end{eqnarray}
where ICF refers to the ``ionization correction factor''.
For comparison, we also plot in Figure 2 the results from the model grid of
Stasi\'nska (1990) for single stars with T$_{eff}$ from 35,000 K to 55,000 K.
Notice that the model points fall within a well-defined band in the diagram,
with the spread in the band increasing as the fraction of O$^{+2}$ increases
(toward the left), as well as the good agreement between our models for OB
clusters and those of Stasi\'nska for single values of T$_{eff}$. The spread
in X(O$^{+2}$) reflects mainly the effects of changing the ionization parameter
in the models. At the same time, the spread in X(C$^{+2}$)/X(O$^{+2}$) reflects
mainly changes in the hardness of the ionizing radiation: for cooler stars (or,
equivalently, an older cluster), the bound-free absorption edges in the stellar
atmospheres are stronger, resulting in fewer photons capable of ionizing C
to C$^{+3}$. Thus, in the case of a nebula with high ionization parameter but
relatively cool stars, essentially all of the C and O are twice ionized.


For the H~II regions we have observed, X(O$^{+2}$) has been measured from
the optical spectra, and are listed in Table 3. Therefore, we can estimate
the correction for unobserved ions of carbon
directly from Figure 2. For our measured values of X(O$^{+2}$), we find the
mean value of X(C$^{+2}$)/X(O$^{+2}$) from the diagram; the correction from
the observed C$^{+2}$/O$^{+2}$ to C/O is then given by equation (4). The
uncertainty in that correction is determined by the spread in
X(C$^{+2}$)/X(O$^{+2}$) at fixed X(O$^{+2}$).
The correction factors (ICFs) derived for our objects are listed
in Table 3 along with their uncertainties, as well as the final values
of the C/O abundance ratios in our objects.

\subsection{Uncertainties}

The dominant sources of uncertainty in our abundance determinations are the
finite S/N in our O~III] line measurements and the uncertainty in the
ionization
correction factor (ICF). The uncertainties in the line strengths listed in
Table 2 were determined from the statistical noise in the spectra over the
full width at zero intensity for each measured line combined in quadrature
with the uncertainty in the relative spectrophotometry of the FOS, as discussed
in \S 2.1. We take the uncertainty in the ICF to be the full range in
X(C$^{+2}$)/X(O$^{+2}$) corresponding to each observed value of X(O$^{+2}$),
as determined from Figure 2. Uncertainties in the electron temperature
contribute to the error in the abundance ratio through the exponential term
in equation (2); the weak dependence of
C$^{+2}$/O$^{+2}$ on T$_e$ means that uncertainties of less than 1000 K in
T$_e$ have only a small effect on the abundance ratio, less than 10\%.
Errors due to uncertainty in the reddening correction are less than 2\% for
all of our objects. Uncertainties in the collision strengths introduce
approximately 8\% uncertainty in the C/O abundance ratios, neglecting
possible systematic errors in the quantum calculations.

There is some concern that the imprecise position of the spectrograph
for I Zw 18 and T1214-277 could lead to errors in physical conditions
and hence in the C/O ratios. The most likely effect of the displacement
is that the data come from regions with slightly lower ionization and/or
electron temperature. Both effects would mean our derived C/O ratios
are artificially enhanced compared to the true values. Lower intrinsic
C/O ratios in these two galaxies would reinforce the trends we observe.

\section{Discussion}

\subsection{Trends in Relative Abundances}

The resulting carbon and oxygen abundances for the targets from our sample are
displayed in Figure 3, plotted as log (C/O) vs. log (O/H). We also plot for
comparison the values for the solar system (Grevesse \& Noels 1993), and the
mean abundances in B stars from Gies \& Lambert (1992; GL) and Cunha \& Lambert
(1993; CL), as well as the abundances for the Orion Nebula (mean of four
studies listed in CL), and NGC 2363 (Peimbert et al. 1986). The latter point
is based on a detection of O~III] 1661-66 \AA\ in both high-dispersion and low
dispersion IUE spectra; note the good agreement between our results for NGC
2363
and those from the IUE data. The following discussion is based solely on those
points for which we have O~III] detections, i.e., excluding the lower limit for
I~Zw~18.


The data show an apparently continuous increase in C/O with increasing O/H in
these metal-poor systems; the behavior of log(C/O) vs log(O/H) is consistent
with a power law with slope 0.43$\pm$0.09 (least-squares fit to the 6 FOS
points, excluding I~Zw~18).
A formal fit gives
\begin{equation}
{log(C/O) = A + B\ log(O/H) },
\end{equation}
where A = 1.07$\pm$0.39 and B = 0.43 $\pm$ 0.09. Note that this fit applies
only to the abundance range log O/H = $-$4.7 to log O/H = $-$3.6. By
comparison,
a simple chemical evolution model with instantaneous recycling predicts C/O =
constant if both C and O are primary elements, or C/O $\propto$ O/H if O is
primary and C secondary. Since only primary sources of C are known to exist,
the trend in Figure 3 suggests that either (or both) the instantaneous
recycling
approximation does not hold for both C and O or that the yield of C varies with
respect to O.

Figure 4 shows the variation of C/N with O/H, with nitrogen abundances obtained
from the sources of optical spectra listed above. There is significantly
greater
scatter in this diagram.
If we exclude the point for I~Zw~18, it is difficult to discern a correlation
between C/N and O/H. If we exclude the Orion and solar system points as well,
then a trend of increasing C/N with O/H is apparent. Such a trend contrasts
with the results for C/N from IUE and optical spectra, compiled and presented
by Pagel (1985), which indicated that C/N is uncorrelated with O/H. A larger
sample of galaxies observed with the FOS is needed to confirm and delineate
the trend more clearly.


Figure 5 presents a comparison of our new results with previous results
found in the literature.  Here we are interested in two questions, (1) do
these new results agree with previous results on C/O measurements of
H~II regions in dwarf galaxies and (2) is the pattern of C/O versus O/H
seen in the dwarf galaxies similar to that seen in our Galaxy?


The top panel of Figure 5 addresses the first question.  Here we have plotted
the points from Figure 3 along with results from IUE observations taken
from the literature. The open circles come from Dufour, Schiffer, \& Shields
(1984; DSS84), the open hexagon represents I Zw 18 as from Dufour, Garnett, \&
Shields (1988; DGS88) as adjusted by Dufour \& Hester (1989), and the open
diamond represents UGC~4483
as reported by Skillman (1991; S91). Points which represent observations of
the same galaxy are connected. The abundances are taken directly from the
literature and are not recalculated from original line strengths (since, in
some cases, the line strengths are not reported). As noted earlier for
NGC~2363,
there is reasonably good agreement between the IUE measurements and the HST
measurements for those objects which have both measurements. Our new results
and those of previous studies clearly indicate a need for new, deep FOS
spectra of I~Zw~18 and H~II regions in the Milky Way and other spiral
galaxies to extend the trend of C/O to metallicities outside the range
of our present sample.

The bottom panel of Figure 5 presents C/O and O/H measurements of Galactic
stars taken from the literature. The data are taken from Sneden, Lambert, \&
Whitaker (1979; SLW79), Clegg, Lambert, \& Tomkin (1981; CLT81), Gies \&
Lambert
(1992: GL92), Kilian (1992; K92),
Tomkin et al. (1992; TLLS92), and Cunha \& Lambert (1994; CL94). There is
considerable scatter in log (C/O) for the disk stars, with an average value
of about $-$0.4. For the halo stars, log (C/O) is fairly constant at $-$0.9
(with a scatter consistent with observational errors). If the trend of
increasing C/O with O/H seen in the H~II regions of dwarf galaxies is
supported by more observations, then a clear difference is emerging between
the abundance pattern seen our Galaxy and the dwarf galaxies. The possible
origins of such differences are discussed in section 4.5.

In the highly-ionized nebulae comprising our sample, the singly-ionized
species (O$^+$, N$^+$, etc.) represent only a small fraction of the
total abundance of the various elements in the gas. This has been a
particularly severe problem for nitrogen since only N$^+$ is observable
in the visual spectrum. Measurements of far-IR fine structure lines of
[N~III] and [O~III] have been made by Lester et al. (1987), among others,
to study the variation of N/O in H~II regions and to compare with optical
measurements of N$^+$/O$^+$. They found that N$^{+2}$/O$^{+2}$ exhibited
a radial gradient across the Galaxy, whereas N$^+$/O$^+$ from optical
spectra did not. Garnett (1990) showed that ionization effects could
be very important in interpreting such trends, and that measurements
of both N$^{+2}$ and N$^+$ are needed to determine N/O with certainty.
Unfortunately, the IR and optical samples of N/O measurements have little
overlap, so the true evolution of N/O is still controversial.

One can also look at the UV lines of N~III] near 1750 \AA\ to measure
N$^{+2}$ in hot H~II regions. The lines are very weak and require long
exposures to obtain a good detection; our exposures were not long enough
to obtain more than upper limits on the N~III] intensities. However, for
three of our objects (N88A, NGC 2363, and C1543+091 -- Table 4), the
2$\sigma$ upper limits on N$^{+2}$ are low enough to indicate that
N$^{+2}$/O$^{+2}$ is not significantly larger than N$^+$/O$^+$ from
optical observations. This is consistent with the photoionization models
for metal-poor nebulae by Garnett (1990), who suggested that optical
measurements of N$^+$ alone can provide reliable measurements of the
nitrogen abundances in such objects. This provides some confidence that
our C/N ratios are reliable. Nevertheless, longer integrations of the
brightest objects are needed to obtain good measurements of N~III]
and confirm this hypothesis.

\subsection{Depletion Onto Grains}

Observations of interstellar extinction and line absorption demonstrate the
need to consider depletion of heavy elements (including carbon and oxygen)
onto grains in order to derive accurate elemental abundances in the ISM.
For our analysis we need to examine how much depletion of carbon and oxygen
is appropriate for our H II regions and the possible effect on the observed
trend in the gas-phase C/O ratio. Much of the discussion below is based on
information from Tielens \& Allamandola (1987), Jenkins (1987), and Mathis
(1990).

Theoretical arguments suggest that there should be little (0.1-0.2 dex)
depletion of O onto grains in the diffuse ISM (Meyer 1985, Tielens \&
Allamandola
1987). Greater depletions are expected from the build-up of icy mantles (mainly
water ice) on grains within dense molecular clouds, but these mantles are
not expected to survive the harsher conditions of the diffuse ISM. On the
other hand, observations of interstellar absorption lines from O show
depletions of approximately 0.4 dex (Jenkins 1987), with no discernible
trend with the cloud density. The amount of depletion inferred, however,
depends on the reference abundances used. Typically, solar abundances
have been used as the reference, yielding the above O depletion. More
recently, Gies \& Lambert (1992; GL) and Cunha \& Lambert (1993; CL) have
determined abundances for B stars in the solar neighborhood and Orion;
these studies indicate that the B-star oxygen abundances are systematically
smaller than in the Sun, and in better agreement with the results from nearby
H~II regions. Sofia, Cardelli, \& Savage (1994) have shown that using the
average B-star abundance for O as the reference standard leads to smaller
derived depletions for O in the local ISM -- approximately 0.2 dex.

The situation for carbon is even more uncertain. There are few reliable
measurements of interstellar carbon lines, because the most easily observed
lines are often highly saturated. Cardelli et al. (1993) and Sofia et al.
(1994) discuss recent measurements of weak interstellar C~II line
measurements with the HST. The results for three sightlines, referenced
to solar abundances, indicate a C depletion of about 0.4 dex also, although
the lower depletion of C toward $\xi$ Per suggests that carbon depletion
may be variable; data for more sightlines are certainly needed to address
this question. The inferred depletions are again strongly dependent on the
choice of reference abundances; a comparison of the interstellar C abundances
with the mean value for B stars from GL would lead one to infer little or no
depletion of carbon (Sofia et al. 1994). Such a result would be difficult
to understand. However, GL noted that the distribution of C abundances
for their sample showed a tail toward low abundances, while the distribution
of N abundances shows a tail toward higher abundances. This suggests that
some of the stars may exhibit the signs of CN processing, and that the
mean C abundance quoted by GL could underestimate the true mean of the
original abundances. Indeed, Cunha \& Lambert (1993) obtain a somewhat
higher mean C abundance for their sample of B stars in Orion. Furthermore,
the most recent compilation of solar abundances (Grevesse \& Noels 1993)
has resulted in a slight downward adjustment of the solar O/H by 0.06 dex.
Using the CL abundances as reference alleviates the problem somewhat,
although the inferred carbon depletion is still relatively small
($\approx$ 0.2 dex).

The uncertainties discussed above make it difficult to determine how to
apply the depletion results to our measured abundances. Direct application
of the above results leads to equal depletions of about 0.35 dex for C and
O (based on solar reference abundances) or 0.15 dex for both (based on B
star reference abundances). Applying these depletions changes the C and O
abundances by equal factors, so the trend observed in Figure 3 would be
unchanged, although the abscissa scale would shift to higher values.
Similarly, the C/N ratios in Figure 4 would all change systematically (N
shows little evidence for depletion in the ISM), but the trend in C/N with
O/H would not change.

Another concern is that depletions may vary with metallicity. Unfortunately,
there is little evidence for or against this possibility. Measurements of
interstellar extinction curves appear to show that the interstellar dust-to-gas
ratio (measured by N$_H$/E(B-V)) varies directly as the carbon abundance, and
that the 2175 \AA\ absorption feature (most commonly attributed to carbon-based
compounds) weakens likewise. This $could$ be interpreted as being consistent
with a constant depletion of carbon with metallicity, but the uncertainties
are large. Measurements of Si/O from our FOS spectra (Dufour et al., in
preparation) also suggest that depletions do not vary significantly with
metallicity. We can examine the effect on the observed
abundance ratios by assuming the maximum possible variation. The worst case is
for the depletions based on the solar reference abundances, i.e., 0.4 dex
depletion for C and O at solar neighborhood metallicities. If the C depletion
does not vary, but the O depletion decreases uniformly to zero for log O/H $<$
$-$4.5, then such variation can account almost entirely for the trend in C/O
in Figure 3, and the true trend would be that C/O is uncorrelated with O/H.
(We reject the possibility that depletions can be $larger$ at low
metallicities.)
On the other hand, Garnett (1990) has shown that the average N/O ratio in
irregular
galaxies is constant with O/H, based on zero depletion of oxygen. Since N is
depleted very little in the ISM, an increasing O depletion with metallicity
would
lead to the conclusion that N/O $decreases$ with metallicity in irregulars, a
result very difficult to understand in the context of either nucleosynthesis or
chemical evolution theory.

Another observation of possible relevance comes from the study of Calzetti,
Kinney, \& Storchi-Bergmann (1994). They have derived extragalactic extinction
curves by comparing stellar continua of highly reddened and lightly reddened
starburst galaxies. Remarkably, they note that their derived extinction laws
show no evidence for the 2175 \AA\ feature (after the galaxy spectra are
corrected for Galactic extinction). Because of the large apertures they used
and the high reddening toward many of their targets, scattering and optical
depth effects could partly account for the Calzetti et al. result. On the other
hand, Fitzpatrick (1985) has shown that stars in the region of 30 Doradus show
weaker 2175 \AA\ features than do stars in the LMC located far from 30 Doradus,
while Rosa \& Benvenuti (1994) find similarly weak 2175 \AA\ features in FOS
spectra of giant H~II regions in M101. If the 2175~\AA\ feature is indeed
attributable to carbon-based grains, these results might indicate that such
grains are destroyed in the energetic environment of a starburst, and thus
that carbon depletions might be small in giant H~II regions.

Given the uncertainties in depletions and the lack of information on their
variation with metallicity, at the present time we choose to assume that grain
depletions will not significantly affect the observed trends of C/O and C/N
with O/H, although the abundance scales may change systematically.

\subsection{Nuclear Reaction Rates and C and O Yields}

$^{16}$O is predominantly a product of $\alpha$-particle captures onto $^{12}$C
during the He burning phase of stellar evolution. The amount of O ejected by a
star depends on its mass -- too small a star leaves its oxygen behind in the
core remnant. Stellar evolution models generally show that 10 M$_{\odot}$
is the smallest star that can produce and eject new oxygen. $^{12}$C is also
produced during He burning through the well-known ``triple-$\alpha$'' reaction.
Theoretical models indicate that carbon can be ejected not only by massive
stars,
but also by intermediate mass stars ($\approx$ 2-8 M$_{\odot}$) through the
convective dredge-up of freshly-synthesized carbon during the asymptotic
giant branch evolution phase of these stars. Measurements of carbon abundances
in planetary nebulae show that some of them have enrichments of carbon,
confirming
that dredge-up does occur in at least some intermediate mass stars.

The relative amounts of C and O produced during He burning are sensitive to
the nuclear reaction rate for the $^{12}C(\alpha,\gamma)^{16}O$ reaction, as
demonstrated by the models of Weaver \& Woosley (1993; hereafter WW93). The
value of the cross section for this reaction has been a source of considerable
uncertainty; details of the controversy are given in Rolfs \& Rodney (1988).
Some attempts have been made to determine the rate empirically. Arnett (1971),
for instance, tried to constrain the reaction rate by comparing the solar
system C/O ratio with the predictions of stellar nucleosynthesis. This kind of
analysis assumed that both C and O are produced in massive stars only and that
the C/O ratio does not evolve with time. Clearly, however, the observations of
C/O in Galactic stars and extragalactic H~II regions show that the C/O ratio
has evolved. WW93 tried a similar approach to the problem,
comparing their results for massive star nucleosynthesis with the solar system
abundance pattern, but $excluding$ carbon in the comparison. They found that a
reaction rate of 1.7$\pm$0.5 times the Caughlan \& Fowler (1988; hereafter
CF88)
rate produced the best match to the solar system abundance distribution for
elements between O and Fe.

Recently, improved measurements of the $^{12}C(\alpha,\gamma)^{16}O$ cross
section have been reported by Buchmann et al. (1993) and Zhao et al. (1993).
The results represent improved agreement with each other compared with
earlier measurements. Nevertheless, there is still significant disagreement
in the measurements that requires resolution. In terms of the commonly used
astrophysical S factor S(E), Buchmann et al. measured a value S$_{E1}$(0.3 MeV)
= 57$\pm$13 keV-b, while Zhao et al. obtained a value S$_{E1}$(0.3 MeV) =
95$\pm$32 keV-b for the same reaction. For comparison, WW93's best
estimate corresponds to S$_{E1}$(0.3 MeV) = 102$\pm$30 keV-b.

Here we attempt an independent estimate of the $^{12}C(\alpha,\gamma)^{16}O$
reaction rate by comparing the stellar nucleosynthesis results of WW93 with
the C/O ratios measured in our most metal-poor galaxies. Figure 6 shows our
observed C/O ratios overlaid with the C/O (by number) ratios obtained by
integrating the WW93 results over the stellar mass range 11-42 M$_{\odot}$
with an IMF slope of $-$1.5 (taken from their Table 7); the calculations
shown are for the case of ``nominal'' semiconvection as defined by WW93. The
horizontal dotted lines show the theoretical integrated C/O ratios obtained
for different values of the $^{12}C(\alpha,\gamma)^{16}O$ rate in units of
the CF88 rate; each line is labeled with the value of the reaction rate used
in the calculation.


Figure 6 shows that the observed C/O ratios in our three most metal-poor
galaxies (not including I Zw 18) are consistent with WW93's estimate for
$^{12}C(\alpha,\gamma)^{16}O$. However, our comparison depends on two
important assumptions:

(1) We assume that the most metal-poor galaxies in our sample have abundance
ratios dominated by nucleosynthesis from massive stars. However, if
intermediate
mass stars have contributed a significant amount of carbon in these galaxies,
our observed C/O would require a larger $^{12}C(\alpha,\gamma)^{16}O$ rate. On
the other hand, the WW93 models cover a mass range of only 11 to 40
M$_{\odot}$;
more massive stars tend to produce relatively more oxygen and so a higher upper
mass limit would drive the predicted C/O ratio down.

(2) We also assume that the stellar initial mass function does not change
significantly with metallicity. WW93 also provide nucleosynthesis results
integrated over an IMF with a slope of $-$2.3; the integrated C/O ratio
for this case is an average of only 11\% larger than for slope $-$1.5,
suggesting that the slope of the IMF does not affect abundance ratios
significantly, within known uncertainties.

This comparison should be viewed with caution because of a number of additional
uncertainties which can affect the results. First of all, we have
only a small number of carbon measurements to work with at this point, and
we can not say with confidence yet that we know either the true variation
of C/O with metallicity or the minimum value of C/O in these galaxies. Second,
the final word on theoretical nucleosynthesis has certainly not been written
yet. Convective treatments and stellar mass loss are still highly uncertain
inputs to stellar evolution models and can have significant effects on the
production of C and O. WW93 illustrate the effects of varying the convective
treatment with a second set of stellar models with a reduced or ``restricted''
semiconvection. These models result in integrated C/O ratios in ejected
material up to twice as large as in the ``nominal'' semiconvection treatment.
The effects of stellar mass loss (Maeder 1992, Woosley, Langer \& Weaver 1993)
on stellar evolution and nucleosynthesis have only recently been explored
extensively; we discuss some of the results below.

It is also apparent that more observations of C/O in metal-poor galaxies will
be needed to clearly delineate the evolution of the C/O ratio and to define
its lower bound. In particular, an improved measurement for I~Zw~18 is a
minimum requirement to anchor the low-abundance end.

\subsection{Chemical Evolution Concerns}

The firm result of this study is that the C/O abundance ratio in
low-metallicity
systems is lower than in the solar neighborhood, by up to a factor of five
compared to the Sun, as illustrated in Figure 3. The low C/O ratios we observe
in our most metal-poor galaxies are consistent with the C/O ratios observed by
Reimers et al. (1992) in Lyman-limit and metal-line absorption systems toward
the
QSO HS1700+6416. In seven absorption systems between redshifts 1.8 and 2.6,
they
measured C/O ratios ranging from 0.1 to 0.2. One lower-redshift system, at z =
1.1572, showed a solar C/O ratio. These results combined with ours clearly show
that the C/O ratio has evolved from low values at early times to the
present-day
``cosmic'' ratio.

We begin our interpretation of the trend in Figure 3 within the framework of
the simple ``closed box'' model of chemical evolution, in which a system
evolves without gas flows into or out of the system, and in which stars
are assumed to have a negligible lifetime, ejecting products of nucleosynthesis
as soon as they are formed (the instantaneous recycling approximation). We
also assume initially that the chemical yields for a generation of stars do
not change with time or composition. Consideration of nucleosynthesis sites
for carbon and oxygen indicate that both C and O should be so-called
``primary''
elements (that is, produced from material that was originally hydrogen or
helium in the star); in such a case, the simple model predicts that C and
O should vary in lockstep with a constant ratio. This prediction is clearly
at odds with the observations.

There are several possible ways to explain the variation in C/O with O/H:

(1) {\it Breakdown of the instantaneous recycling approximation.} Stars do
in fact have non-negligible lifetimes. As a result, a temporal variation
in the ratio of two elements will be observed if the two elements are
produced in stars with different lifetimes (Tinsley 1979). It was noted
earlier that carbon can be produced in both massive stars and intermediate
mass stars. If intermediate mass stars dominate in the production of carbon,
then the trend in C/O seen in Figure 3 could be interpreted as an ``age''
effect: the most metal-poor objects would be the youngest, displaying an
enrichment pattern dominated by massive stars only, while the more
metal-rich galaxies show increasing contributions from longer-lived
intermediate mass stars.

(2) {\it Variable yields.} Recent stellar evolution models including mass
loss via winds predict much larger yields of C relative to O (Maeder 1992,
Woosley et al. 1993). If stellar winds are primarily radiatively driven,
and dependent on opacity, then the stellar mass loss rates should depend
on metallicity. Maeder (1992) presents an analysis of the expected yields
from stellar evolution models which incorporate metallicity-dependent
stellar mass loss. These models result in an increasing yield of carbon,
and a decreasing yield of oxygen, with metallicity, and thus, C/O increasing
with O/H. Carigi (1994) has recently calculated
chemical evolution models for the Galaxy which take the Maeder yields
for massive stars into account. Her models show a significantly steeper
increase in C/O with O/H than the older models of Matteucci \& Fran\c cois
(1989), which used massive star yields from Woosley (1986). We plot the
results for Carigi's models 1a and 2 against our data in Figure 7. The
models appear to reproduce the trend in the data fairly well, although
there is a small systematic offset between the models and the data; this
could indicate that the stellar yields require some adjustment, or that
the observed carbon abundances are affected by grain depletion. Prantzos,
Vangioni-Flam, \& Chauveau (1994) have computed similar models for the
solar neighborhood also, and find similar results.


One should use caution in using such models to interpret the observations
at this time. Carbon yields are still relatively uncertain.  The yields
from massive stars are subject to uncertainties in convective mixing,
nuclear reaction rates, and stellar mass loss rates. Theoretical models
suggest metallicity-dependent mass loss is important; observationally
the situation is less certain. Direct measurements of stellar mass-loss
rates have not been precise enough to demonstrate a correlation with Z,
although indirect evidence from the frequency of Wolf-Rayet stars vs.
metallicity is more compelling (Leitherer 1993). Carbon yields for
intermediate mass stars, most commonly taken from Renzini \& Voli (1981),
have not been recomputed with the more recent estimates of the
$^{12}C(\alpha,\gamma)^{16}O$ reaction
rate, and so they are uncertain as well.

(3) {\it Variable IMF.} A variable stellar initial mass function has been
invoked to explain some properties of giant H~II regions and starburst
galaxies (e. g., Melnick, Terlevich, \& Eggleton 1985). If the IMF is
weighted toward more massive stars at low metallicities, then an increasing
C/O ratio with metallicity can be explained as a change in the relative
contributions of massive and intermediate mass stars. Observations of
luminous stars in the Local Group show no convincing evidence for such
a systematic variation in either the slope of the IMF or in the upper
mass limit for stars (Garmany 1989, Parker \& Garmany 1993). Moreover,
varying the IMF affects the abundance ratios of a number of elements,
as illustrated by models for sulfur presented in Garnett (1989). As a
result, variable IMFs are not attractive for explaining variations in
the composition of galaxies.

A comparison of Figures 3 and 4 proves interesting. Note in Figure 4 the
abrupt drop in C/N between the solar neighborhood objects and the more
metal-rich dwarf galaxies (NGC 2363 and LMC), attributable to much larger
nitrogen abundances in the Galactic nebulae. In contrast, the C/O ratio
maintains a smooth upward trend including the Galactic objects. Taken
together, the features in Figs. 3 and 4 suggest that the nucleosynthesis
of C may be largely decoupled from that of N. One possibility is that C
and O production is dominated by massive stars, while N comes mainly from
intermediate mass stars. Another possibility is that N and C both produced
mainly in intermediate mass stars, but in different mass ranges. The
fluctuations in N/O could result from local pollution by Wolf-Rayet stars
rather than delayed ejection of N from IMS (Pagel, Terlevich, \& Melnick
1986, Pagel et al. 1992). At present it is not possible to design a unique
model to account for the observed trends. The true picture will have
repercussions for the use of $\Delta$Y/$\Delta$(element) correlations
to determine the primordial He fraction (Steigman et al. 1989).

\section{Summary}

We have presented UV spectroscopy of H~II regions in low-luminosity dwarf
galaxies and the Magellanic Clouds obtained with the Faint Object Spectrograph
on the Hubble Space Telescope. We have measured O~III] 1666 \AA\ and C~III]
1909 \AA\ simultaneously, enabling us to determine C$^{+2}$/O$^{+2}$ with
greatly reduced uncertainties from errors in electron temperature and
reddening. These FOS spectra represent a significant improvement over IUE
in terms of sensitivity and reliability of abundance measurements. The
results indicate that further FOS spectroscopy of H~II regions will
greatly improve our understanding of H~II regions and the composition
of the interstellar medium.

The results from our HST observations show a monotonic increase in log (C/O)
from $\sim$ $-$0.9 at log (O/H) $\approx$ $-$4.6 to $\sim$ $-$0.5 at log (O/H)
$\approx$ $-$3.7. This trend is different from that seen in the Galactic stars,
where C/O is roughly constant at log (C/O) = $-$0.9 over the same range in O/H.
One possible interpretation of this trend is that the most metal-poor galaxies
are the youngest and dominated by the products of early enrichment by massive
stars, while more metal-rich galaxies show increasing, delayed contributions of
carbon from intermediate mass stars. However, recent evolution models for
massive
stars including mass loss from stellar winds suggest that the yield of carbon
from massive stars may increase with metallicity relative to the yield of
oxygen,
complicating the interpretation of the abundance trends.

The observed C/O ratios in our three most metal-poor galaxies are consistent
with the estimate of Weaver \& Woosley (1993) for the
$^{12}C(\alpha,\gamma)^{16}O$.
cross-section. This result is subject to uncertainty from the poorly-known
contribution of carbon from intermediate mass stars in these galaxies. Improved
nucleosynthesis calculations for intermediate mass stars are needed to enhance
the interpretation of the observed trend in C/O.

The abrupt drop in C/N between 30 Dor and the solar neighborhood (Fig. 4) is
intriguing. The result suggests that most of the nitrogen production in our
Galaxy is delayed with respect to both carbon and oxygen, in contrast with
previous results indicating that C and N varied in lockstep (Pagel 1985). It
may be that Fig. 4 indicates that spirals and irregulars experience different
chemical evolution histories, and that one cannot safely combine abundance
data from both irregulars and spirals to study the abundance evolution of
every element. Certainly, more FOS observations of carbon are needed to
confirm the trend in Fig. 4, particularly measurements in H~II regions in
spiral galaxies.

Another question not addressed here is that of the intrinsic dispersion in
C/O. It has been suggested often (e. g., Clayton \& Pantelaki 1993, Pilyugin
1992, 1993) that dwarf galaxies experience intermittent starbursts, and that as
a result the delayed ejection of fresh N and C from intermediate mass stars
should lead to relatively large fluctuations in N/O and C/O at fixed O/H.
The data in Fig. 3 suggest only a small intrinsic dispersion in C/O, but
the sample is small, the uncertainties are still relatively large, and the
target selection may have been fortuitous. FOS measurements of C/O for a
larger sample of dwarf galaxies would permit us to address the dispersion
in C/O with greater confidence.

\acknowledgments

This study has benefitted from conversations with R. Bohlin, J. Cardelli,
W. D. Arnett, K. Nomoto, and J. Mathis. We thank Leticia Carigi for providing
us with machine-readable tables of her chemical evolution model results, and
Bernard Pagel for helpful comments on the manuscript. Support for this work
was provided by NASA through grant GO-3840.01-91A from the Space Telescope
Science Institute, which is
operated by the Association of Universities for Research in Astronomy, Inc.,
for NASA under NAS5-26555. Support for DRG is provided by NASA and STScI
through the Hubble Fellowship award HF-1030.01-92A. EDS acknowledges support
from NASA-LTSARP grant NAGW-3189. EDS, ET, and RJT acknowledge support from
a NATO grant for Collaborative Research, No. CRG910269.

\clearpage


\begin{planotable}{lccccc}
\tablewidth{30pc}
\tablecaption{Journal of FOS Observations}
\tablehead{
\colhead{Object} & \colhead{Observation ID} & \colhead{ } &
\colhead{Integration Time} & \colhead{ } & \colhead{Aperture} }
\startdata
30 Doradus  & Y1470202T & & 1800s & & B-3 \nl
            & Y1470203T & & 1800s & & B-3 \nl
            &           & &       & &     \nl
C1543+091   & Y1470802T & & 1800s & & B-3 \nl
            & Y1470803T & & 1800s & & B-3 \nl
            & Y1470804T & & 1800s & & B-3 \nl
            &           & &       & &     \nl
T1214-277   & Y1470702T & & 1800s & & B-3 \nl
            & Y1470703T & & 1800s & & B-3 \nl
            & Y1470704T & & 1800s & & B-3 \nl
            &           & &       & &     \nl
NGC 2363    & Y1LL0102T & & 2100s & & C-1 \nl
            & Y1LL0103T & & 1200s & & C-1 \nl
            &           & &       & &     \nl
SBS0335-052 & Y1470102T & & 1800s & & B-3 \nl
            & Y1470103T & & 1800s & & B-3 \nl
            & Y1470104T & & 1800s & & B-3 \nl
            &           & &       & &     \nl
I Zw 18     & Y1470502M & & 1650s & & B-3 \nl
            & Y1470503M & & 1650s & & B-3 \nl
            &           & &       & &     \nl
SMC N88A    & Y1M70302T & & 1200s & & C-4 \nl
\end{planotable}
\clearpage
\begin{planotable}{lccc}
\tablewidth{0pc}
\tablecaption{Line Intensities from FOS Observations.}
\tablehead{
\colhead{Object} & \colhead{O III]} & \colhead{N III]} & \colhead{C III]}
\\[.2ex]
\colhead{} & \colhead{$\lambda$1666} & \colhead{$\lambda$1750} &
\colhead{$\lambda$1909} }
\startdata
\cutinhead{Fluxes in 10$^{-15}$ ergs cm$^{-2}$ s$^{-1}$}
30 Doradus  &     1.22$\pm$0.95 & $<$1.3 &  12.05$\pm$0.32  \nl
C1543+091   &     3.35$\pm$0.73 & $<$1.3 &   9.80$\pm$0.32  \nl
T1214-277   &     0.46$\pm$0.35 & $<$0.5 &   1.18$\pm$0.13  \nl
NGC 2363    &    13.3$\pm$2.2   & $<$5.1 &  57.7$\pm$1.3    \nl
SBS0335-052 &     3.22$\pm$0.98 & $<$1.6 &   5.79$\pm$0.47  \nl
I Zw 18     & $<$ 2.9           & $<$1.7 &   2.40$\pm$0.41  \nl
SMC N88A    &    19.5$\pm$1.4   & 3.2$\pm$2.1 & 106.0$\pm$5.6    \nl
\cutinhead{Line Strengths Normalized to I($\lambda$1909)}
30 Doradus  & 0.10$\pm$0.08 & $<$0.01 & 1.00  \nl
C1543+091   & 0.34$\pm$0.08 & $<$0.13 & 1.00  \nl
T1214-277   & 0.39$\pm$0.31 & $<$0.40 & 1.00  \nl
NGC 2363    & 0.23$\pm$0.04 & $<$0.09 & 1.00  \nl
SBS0335-052 & 0.56$\pm$0.18 & $<$0.28 & 1.00  \nl
I Zw 18     & $<$ 1.2       & $<$0.70 & 1.00  \nl
SMC N88A    & 0.27$\pm$0.03 & 0.03$\pm$0.02 & 1.00  \nl
\end{planotable}
\clearpage
\special{pslandscape}
\begin{planotable}{lccccccc}
\tablewidth{0pc}
\tablecaption{C/O Abundance Ratios from FOS Observations}
\tablehead{
\colhead{Object} & \colhead{30 Doradus} & \colhead{SMC N88A} & \colhead{NGC
2363}
& \colhead{C1543+091}& \colhead{T1214-277} & \colhead{SBS0335-052} &\colhead{I
Zw 18} }
\startdata
Log O/H & $-$3.70$\pm$0.10 & $-$3.91$\pm$0.04 & $-$4.08$\pm$0.04 &
$-$4.24$\pm$0.10 & $-$4.41$\pm$0.05 & $-$4.64$\pm$0.06 & $-$4.83$\pm$0.04 \nl
T$_e$   & 10,400$\pm$400 & 14,000$\pm$500 & 14,800$\pm$500 & 16,100$\pm$500 &
17,800$\pm$800 & 19,500$\pm$800 & 19,600$\pm$900 \nl
$C^{+2} \over O^{+2}$ & 0.31$\pm$0.25 & 0.15$\pm$0.02 & 0.18$\pm$0.04 &
0.13$\pm$0.04 & 0.12$\pm$0.10 & 0.087$\pm$0.035 & $>$ 0.04 \nl
X($O^{+2}$) & 0.83 & 0.96 & 0.94 & 0.90 & 0.95 & 0.91 & 0.84 \nl
ICF & 1.06$\pm$0.14 & 1.33$\pm$0.48& 1.31$\pm$0.45 & 1.20$\pm$0.30 &
1.32$\pm$0.45 & 1.21$\pm$0.31 & 1.09$\pm$0.15 \nl
Log C/O & $-$0.48$\pm$0.26 & $-$0.72$\pm$0.17 & $-$0.63$\pm$0.15 &
$-$0.81$\pm$0.15 & $-$0.80$\pm$0.28 & $-$0.94$\pm$0.17 & $>$-1.35 \nl
$N^{+2} \over O^{+2}$ & $<$ 0.10 & 0.016$\pm$0.011 & $<$ 0.04 & $<$0.04 & $<$
0.12 & $<$ 0.06 & \nodata   \nl
\end{planotable}
\clearpage
\special{psportrait}
\begin{planotable}{lccccr}
\tablewidth{16pc}
\tablecaption{Adopted O III $^5$S-$^3$P Collision Strength}
\tablehead{
\colhead{ } & \colhead{T$_e$ (K)} & \colhead{ } &
\colhead{ } & \colhead{$\Omega$($^5$S-$^3$P)} &\colhead{ } }
\startdata
 & 10,000 & & & 1.235 &  \nl
 & 15,000 & & & 1.284 &  \nl
 & 20,000 & & & 1.295 &  \nl
\end{planotable}

\clearpage

\begin{figure}
\caption{FOS spectra of two giant extragalactic H~II regions, showing the
O~III] 1666 \AA, Si III] 1883, 1892 \AA, and C~III] 1908 \AA\ features.
Top: 30 Doradus. Bottom: SBS0335-052. Both spectra have been smoothed with
a three-point Gaussian filter.}
\end{figure}

\begin{figure}
\caption{The ionization of C$^{+2}$ and O$^{+2}$ in H~II region models, as
described in the text. X(C$^{+2}$ and X(O$^{+2}$) are the fractions of
C and O in the doubly-ionized state. (X(C$^{+2}$/X(O$^{+2}$))$^{-1}$ is
the correction (ICF) to convert C$^{+2}$/O$^{+2}$ into C/O.}
\end{figure}

\begin{figure}
\caption{The C/O abundance ratio in irregular galaxies vs. O/H. Filled circles
are the data from our FOS spectra; unfilled squares are data for Orion and
NGC 2363 obtained from
IUE spectra. The stars represent the mean abundances in B stars determined by
Gies \& Lambert (1992; GL) and Cunha \& Lambert (1993; CL).}
\end{figure}

\begin{figure}
\caption{C/N abundance ratio vs. O/H in irregular galaxies. Symbols are the
same as in Figure 3.}
\end{figure}

\begin{figure}
\caption{The C/O abundance ratio versus O/H for H~II regions in irregular
galaxies and Galactic stars.  In the top figure the new HST results are
plotted with previous results from IUE observations. In the lower figure
observations of Galactic stars are presented. The symbols are described
in the text.}
\end{figure}

\begin{figure}
\caption{Comparison of our C/O abundance ratios in irregular galaxies with
the results of stellar nucleosynthesis calculations (Weaver \& Woosley 1993).
The horizontal dotted lines indicate the value of the C/O ratio one obtains
by integrating the C and O produced by massive stars over a mass range 11 - 42
M$_{\odot}$ and an IMF with slope $-$1.5, for various values of the
$^{12}C(\alpha,\gamma)^{16}O$ reaction rate. Each line is labeled with the
magnitude of the reaction rate, as defined in WW93; the reaction rate increases
from 0.5 times the CF88 rate for the top line to 3.0 times CF88 for the
bottom line.}
\end{figure}

\begin{figure}
\caption{Comparison of C/O abundances in irregular galaxies with chemical
evolution models from Carigi (1994) and Matteucci \& Fran\c cois (1989);
two of Carigi's models are represented by the solid and dotted lines; model
1a represents evolution for a Salpeter IMF, while model 2 is for a Scalo
IMF. Both models use massive star yields from Maeder (1992). The Matteucci
\& Fran\c cois model is shown by the dot-dash line; this model uses massive
star yields from Woosley (1986). Note that the chemical evolution models
were designed to simulate the evolution of the solar neighborhood.}
\end{figure}

\end{document}